\documentclass[epj,final]{svjour}
\usepackage{epsfig}
\usepackage{amsmath,amssymb}
\begin{document}
%%%%%%%%%%%%%%%%%%%%%%%%%%%%%%%%%%%%%%%%%%%%%%%%%%%%%%%%%%%%%%%%%%%%%%%%%%%
\newcommand{\vm}{v_{\text{max}}}
\newcommand{\vs}{v_{\text{safe}}}
\newcommand{\vd}{v_{\text{des}}}
\newcommand{\vi}{v_{\text{init}}}
\newcommand{\vc}{v_{\text{crit}}}
\newcommand{\gi}{g_{\text{init}}}
\newcommand{\lc}{l_{\text{car}}}
\newcommand{\q}{\bar{p}}
\newcommand{\btau}{\boldsymbol{\tau}}
\newcommand{\bttau}{\boldsymbol{\tilde{\tau}}}
\newcommand{\btaujn}{\boldsymbol{\tau}_j^{(n)}}
\newcommand{\balpha}{\boldsymbol{\alpha}}
\newcommand{\bbeta}{\boldsymbol{\beta}}
\newcommand{\nn}{{\cal N}}
\newcommand{\Pt}{\tilde{\mathbf P}}
\newcommand{\cP}{{\cal P}}
\newcommand{\be}{\begin{equation}}
\newcommand{\ee}{\end{equation}}
\newcommand{\bea}{\begin{eqnarray}}
\newcommand{\eea}{\end{eqnarray}}
\newcommand{\nonu}{\nonumber\\}
%%%%%%%%%%%%%%%%%%%%%%%%%%%%%%%%%%%%%%%%%%%%%%%%%%%%%%%%%%%%%%%%%%%%%%%%%%%%%
\title{\bf Boundary-induced phase transitions in a space-continuous
traffic model with non-unique flow-density relation}

\author{Alireza Namazi\inst{1,2} \and Nils Eissfeldt\inst{2} \and
Peter Wagner\inst{3} \and Andreas Schadschneider\inst{1}}
\institute{Institut f\"ur Theoretische Physik, Universit\"at zu K\"oln,
D--50937 K\"oln, Germany
\and
ZAIK -- Center for Applied Informatics, Universit\"at zu K\"oln, 
D-50931 K\"oln, Germany   
\and
Institute for Transportation Research, German Aerospace Center, D--12489
Berlin, Germany
}

\authorrunning{A.\ Namazi et al.}
\titlerunning{Boundary-induced phase transitions in a space-continuous
traffic model}
\date{\today}

%%%%%%%%%%%%%%%%%%%%%%%%%%%%%%%%%%%%%%%%%%%%%%%%%%%%%%%%%%%%%%%%%%%%%
%%%%%ABSTRACT
%%%%%%%%%%%%%%%%%%%%%%%%%%%%%%%%%%%%%%%%%%%%%%%%%%%%%%%%%%%%%%%%%%%%%
\abstract{
  The Krauss-model is a stochastic model for traffic flow
  which is continuous in space.
%, collision-free , discrete in time and has bounded
%  acceleration and deceleration. 
  For periodic boundary conditions it
  is well understood and known to display a non-unique flow-density
  relation (fundamental diagram) for certain densities. 
  In many applications, however, the behaviour under open boundary
  conditions plays a crucial role.
% since such systems show boundary-induced phase transitions. 
  In contrast to all models investigated so far, the high flow states 
  of the Krauss-model are not metastable, but also stable. 
%  Therefore fluctuations intrinsic to the model are not able to destroy these 
%  system states.
%  It is shown that also in such a state-continuous model with a
%  bistable fundamental diagram 
  Nevertheless we find that the current in open systems obeys an extremal 
  principle introduced for the case of simpler discrete models. 
  The phase diagram of the open system will be completely determined by
  the fundamental diagram of the periodic system through this
  principle. In order to allow the investigation of the whole state
  space of the Krauss-model, appropriate strategies for the injection of 
  cars into the system are needed. 
  Two methods solving this problem are discussed and the boundary-induced 
  phase transitions for both methods are studied. 
  We also suggest a supplementary rule for the extremal principle to
  account for cases where not all the possible bulk states are generated
  by the chosen boundary conditions.}

%{05.60.-k}{Transport processes}}

\maketitle

%%%%%%%%%%%%%%%%%%%%%%%%%%%%%%%%%%%%%%%%%%%%%%%%%%%%%%%%%%%%%%%%%%%%%
%%%%%%%%%%%%%%%%%%%%%%%%%%%%%%%%%%%%%%%%%%%%%%%%%%%%%%%%%%%%%%%%%%%%%
%%%%%%%%%%%%%%%%%%%%%%%%%%%%%%%%%%%%%%%%%%%%%%%%%%%%%%%%%%%%%%%%%%%%%
%%%%%%%%%%%%%%%%%%%%%%%%%%%%%%%%%%%%%%%%%%%%%%%%%%%%%%%%%%%%%%%%%%%%%
%%
\section{Krauss-model}
\label{sec_SK_Model}

The number of vehicles on highways and in cities is increasing each
year causing vehicular traffic to suffer more and more from jams.
The phenomena related to traffic jams have attracted the attention of
physicists and engineers since almost half a century, trying to
develop models describing the features of the real traffic.
Generally there are two different approaches: microscopic and
macroscopic \cite{shadRep,Helrev,nagatani}. 
Whereas in microscopic models different
vehicles and their dynamics can be distinguished, in macroscopic models
only densities are considered, similar to hydrodynamcis.

However, the approach of a physicist is usually quite different from
that of a traffic engineer. One of the current interests of
statistical physicists are the so called "nonequilibrium systems".
In microscopic vehicular traffic theories, vehicular
traffic is treated as a system of interacting particles driven far
from equilibrium and offers the possibility to study various
fundamental aspects of the dynamics of truly nonequilibrium systems.

Empirical observations show that the average velocity decreases with 
increasing vehicle density. So the average current (or flow), which is 
the product of average velocity and density, is a function of the density.
%and can be shown as a curve in the density-current plane. 
The functional relation between current and density is usually called 
fundamental diagram. Its generic form can be understood easily. 
For small densities all vehicles can move with their desired velocity $\vm$ 
and the current increases monotonously. For large densities the vehicles 
interact with each other and the average velocity is much smaller than 
in the free flow regime. This causes a decrease of the current, with a 
maximum at an intermediate value.

The traffic model introduced in \cite{krauss,skthes,krauss2}, called
Krauss-model in the following, is based on an approach by Gipps
\cite{gipps} considering the braking distance of individual cars. 
Starting from the assumption of safe driving an update scheme can be 
formulated in the manner of the well-known Nagel-Schreckenberg (NaSch) 
model \cite{nasch,Schr:Scha:Na:Ito}. In the Krauss-model --- unlike the 
NaSch model --- the state variables, i.e.\ space and velocity, are chosen 
to be continuous. To make the model safe, i.e.\ free of collisions, a 
safe velocity $\vs$ for each car is introduced, which is calculated in 
every timestep taking into account that there is a maximum
acceleration and deceleration rate for each car. The vehicles will be
updated in parallel corresponding to discrete time dynamics.

\begin{figure}[t]
  \centerline{\epsfig{figure=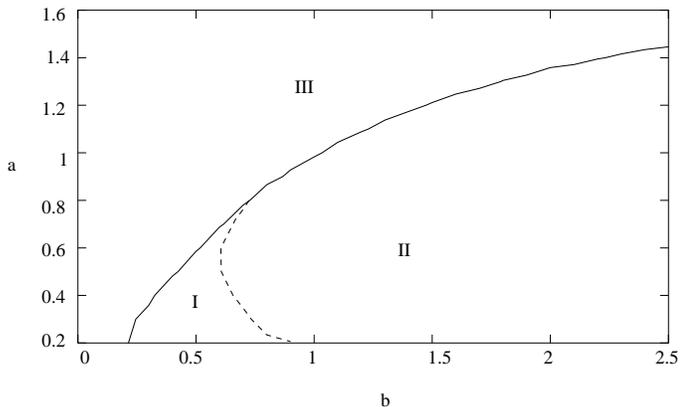,width=9cm}}
  \caption{Classes of qualitatively different behaviour in the Krauss-model. 
    In class $I$, the acceleration $a$ and deceleration $b$ are realistic 
    and all properties $1-4$ of jams are present. In class $II$
    decelerations are large and the properties $1$,$3$, and $4$ of
    jams are not reproduced. In class $III$ accelerations are
    large and no jams exist (from \cite{skthes}).}
  \label{fig:types}
\end{figure}

The model has been designed to reproduce the empirial findings in
traffic jams \cite{kerner1,kerner2,kerner3}:
\begin{enumerate}  
\item There is a density regime with non-unique flow-density relation.
\item Traffic jams can develop and exist under ``pure'' conditions, 
  i.e.\ in the absence of any obstacles. 
\item The flux out of a jam is not maximal. 
\item The outflow from jams is stable.
\item The outflow from jams and the velocity of the downstream front do 
    not depend on the inflow conditions.
\end{enumerate}
These properties are displayed by the model for a certain range of
parameters. The model equations proposed in \cite{skthes} even show a
much richer behaviour depending on the relation between the
ac--  and deceleration capabilities \cite{skthes,krauss2}. Three
different domains can be distinguished (see Fig.~\ref{fig:types}):
\begin{itemize}
\item Class $I$:
  \begin{itemize}
  \item Accelerations and decelerations are realistic and bounded. 
  \item All properties of jams are modeled correctly. 
  \item The jamming transition is a first order phase transition.
  \item The interactions are effectively long ranged. 
  \end{itemize} 
\item Class $II$:
  \begin{itemize}
  \item Decelerations are unbounded, leading to effectively short 
    ranged interactions.
  \item Properties $1$, $3$, and $4$ of jams are not reproduced.
  \item The jamming transition is no phase transition, but a crossover.
  \end{itemize}
\item Class $III$:
  \begin{itemize}
  \item Accelerations are unbounded.
  \item No structure formation at all.
  \end{itemize}
\end{itemize}
Throughout this article only models of class $I$ will be investigated, 
i.e.\ stable jams can occur and the flow-density relation is 
not unique in a certain density regime (Fig.~\ref{fig:FDG_p}).
The stability of the jams is directly related to the fact that the
outflow from a jam is smaller than the maximal possible flow
\cite{localcluster}.

Recently an alternative classification of stochastic traffic models 
with respect to two properties has been suggested \cite{NKW}: 
\begin{itemize}
\item the stability of the high-flow states.  
\item the stability of the outflow interface of jams.
\end{itemize}
These stability criteria were introduced to obtain a clearer
characterization of traffic flow models with the focus on their
stochastic properties. The Krauss-model of class $I$ exhibits
stable high-flow states and a stable jam interface \cite{NKW}. 
Stability of high-flow states means that the intrinsic stochasticity
of the model is not sufficient to cause a transition into the jammed
regime. Therefore, the dynamics of the Krauss-models differs from the
VDR-model \cite{VDR,newVDR}, for which an unstable interface was
found. For the latter, the high-flow states are truly metastable,
i.e.\ for increasing system length the probability for a transition into
the jammed state becomes equal to one. In order to emphasize the
difference between the nature of the high--flow states in the two
models the term {\em bistable} is used in the context of the Krauss-model.

%%%%%%%%%%%%%%%%%%%%%%%%%%%%%%%%%%%%%%%%%%%%%%%%%%%%%%%%%%%%%%%%%%%%%
%%%%%%%%%%%%%%%%%%%%%%%%%%%%%%%%%%%%%%%%%%%%%%%%%%%%%%%%%%%%%%%%%%%%%
%%
\subsection{Dynamical equations}
To derive the underlying dynamical equations, two types of motion of
vehicles are considered. The first type is free motion, the second 
the motion of a vehicle while interaction with
another vehicle takes place. Corresponding to this, two main
assumptions can be made. The free motion is bounded by some maximum 
velocity $\vm$: 
\begin{equation}
  v \leq \vm\; .
  \label{defvm}
\end{equation}
It is assumed that the system remains free of collisions
and that a driver always chooses a velocity that does not exceed
the maximum safe velocity $\vs$ which guarantees the absence of
collisions:
\begin{equation}
  v \leq \vs\; .
\label{defvs}
\end{equation}
$\vs$ is determined from the condition that the braking distance
$d(v)$ needed to stop when moving with velocity $v$ satisfies
\begin{equation}\label{eq:safety_gipps}
  d(v_f) + v_f\tau \leq d(v_l) + g\, .
\end{equation}  
The quantity on the left side is the braking distance of the following 
car (velocity $v_f$) including a finite reaction time $\tau$. This 
distance has to be smaller than the braking distance of the leading
car (moving with velocity $v_l$) plus the gap $g$ between the
vehicles. Furthermore the model takes into account that positive and
negative accelerations are bounded:
\begin{equation}
  -b \leq \frac{d v}{d t} \leq a\; , \qquad \text{with\ \ }   a,b > 0.
\end{equation} 
It is natural to implement these restrictions using a continuous space 
variable, but time is discrete with timesteps $\Delta t$. From the
above restrictions the dynamics of the model can be derived.
    
It will be assumed that, apart from random fluctuations, every vehicle
moves at the highest velocity compatible with the restrictions stated
above. In this way the model can be formulated immediately, giving
\begin{equation}\label{eq:THE model}
  \begin{array}{lcl}
    \vs(t) &=& v_l(t)+\frac{g-v_l(t)}{
      \frac{v_f(t)+v_l(t)}{2b}+\tau}\; ,\\[2ex]
    \vd(t) &=& \min\{\vm, v(t) + a\Delta t, \vs(t)\}
    \; ,\\[2ex]
    v(t+\Delta t) &=& \max\{0, v_{\rm des}(t) -\eta\}\; ,\\[2ex]
    x(t+\Delta t) &=& x(t) + v \Delta t\; .
  \end{array}
\end{equation}  
Here the gap $g = x_{l}- x_{f} - \lc$ is the spatial headway between the
leading car at $x_l$ and the following car at $x_f$, where $\lc$ denotes the
length of a car. $v_l$ and $v_f$ are the velocities of leading and
following cars, respectively. The safe velocity $\vs$ has to be
determined in accordance with condition (\ref{eq:safety_gipps}).
%that the braking distance of the following car is smaller than the gap to 
%the leading car where also a finite reaction of the driver has to be taken 
%into account. 
$\vd$ is the desired velocity representing the wish to drive as fast
as possible through the acceleration $v+ a\Delta t$, but also
respecting the conditions (\ref{defvm}) and (\ref{defvs}). The random
perturbation $\eta > 0$ has been introduced to allow for deviations
from optimal driving, where $\eta = \epsilon \xi$ and $\xi$ is a
random number uniformly distributed in the interval $[0,1]$. 

In the following we set $\Delta t=\tau =1$.
The other generic parameters used in the simulations are 
\begin{equation}\label{eq:params}
  \begin{array}{lcl}
    a= 0.1 ,\  b= 0.6 ,\  \vm= 5 ,\  \epsilon= 1.0 ,\  \lc= 1.0\ . 
  \end{array}
\end{equation} 
The unit of the space coordinates is the length $\lc$ of one car.
Another parameter is the length $L$ of the system which has been
choosen to be equal to $2001$ (if not stated otherwise).

%%%%%%%%%%%%%%%%%%%%%%%%%%%%%%%%%%%%%%%%%%%%%%%%%%%%%%%%%%%%%%%%%%%%%
%%%%%%%%%%%%%%%%%%%%%%%%%%%%%%%%%%%%%%%%%%%%%%%%%%%%%%%%%%%%%%%%%%%%%
%%
\subsection{Characteristics of the model}

\begin{figure}[t]
  \centerline{\epsfig{figure=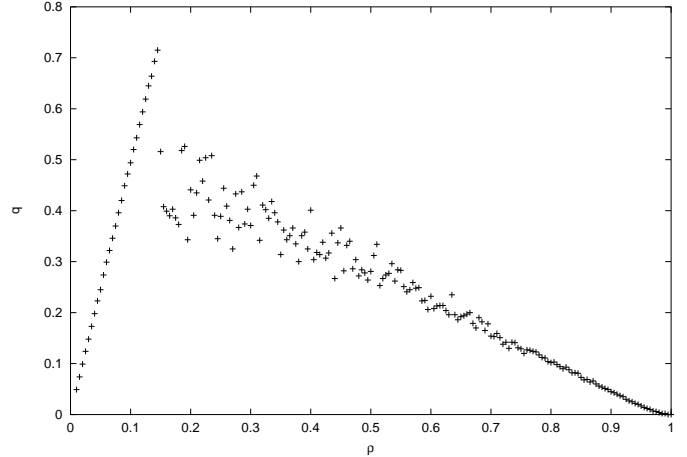,width=9cm}}
  \caption{Fundamental diagram of the Krauss-model with periodic boundary
    conditions and $a=0.1$, $b= 0.6$, $\vm= 5$, $\epsilon= 1.0$,
    $\lc= 1.0$. The density is a mean value of densities measured in
    an interval of length $\frac{L}{3}$ located in the middle of a
    system of length $L = 2001$. The flow $q$ is measured locally
    in the middle of the system. For densities $ 0.1 \le \rho \le
    0.14$ the flow is {\it bistable} such that a stable high flow 
    branch can be observed.}
  \label{fig:FDG_p}
\end{figure}

For the parameters chosen in (\ref{eq:params}) the Krauss-model
belongs to class $I$ and exhibits a bistable region with a stable high
flow branch which implies a non-unique flow-density relation.  
Fig.~\ref{fig:FDG_p} shows a fundamental diagram for a system 
corresponding to class $I$.

The existence of a bistable regime is related to the occurance of phase
separation in the system. The distribution of the gaps and velocities
of a system in the jammed state has two peaks \cite{skthes}, i.e.\
there are two groups of cars in the system.  It separates
into a macroscopic jam and a free-flow region. According to the
initial conditions there exists another system state in which cars
drive with velocities close to $v_{\rm max}$ and the distribution of
gaps possesses only one peak. These states belong to the high--flow
branch in the ambiguous part of the fundamental diagram.

%%%%%%%%%%%%%%%%%%%%%%%%%%%%%%%%%%%%%%%%%%%%%%%%%%%%%%%%%%%%%%%%%%%%%
%%%%%%%%%%%%%%%%%%%%%%%%%%%%%%%%%%%%%%%%%%%%%%%%%%%%%%%%%%%%%%%%%%%%%
%%
\section{Open boundary conditions}
\label{sec_ob}
%%
%%%%%%%%%%%%%%%%%%%%%%%%%%%%%%%%%%%%%%%%%%%%%%%%%%%%%%%%%%%%%%%%%%%%%
%%
%%\subsection{Introduction}
%%\label{ss_ob_intro}   
%%
One of the most significant differences between systems with open and
periodic boundary conditions is the car density $\rho$, which in a
periodic system is a conserved quantity. Here the density and the 
initial conditions (in the bistable regime) determine the stationary
state completely, which allows to study the density-dependence of the
macroscopic parameters. In systems with open boundary conditions (OBC) one
has to deal with two different tuning parameters, namely the injection
rate $\alpha$ and the extraction rate $\beta$. So the car density in
the bulk will be a result of these rates and the underlying model
dynamics. In general a nontrivial density profile will develop, i.e.\
the average density in the system will depend on the position.

The influence of $\alpha$ and $\beta$ on the car density implies that
quantities like bulk density\footnote{Measured by averaging over an 
interval of length $\frac{L}{3}$ in the middle of a system of length
$L$.}, current (flow) $q$, and the density profiles show a different behaviour 
than in periodic systems, which were studied extensively for cellular 
automata, optimal velocity models etc.
(see e.g.\ \cite{shadRep,Helrev,nagatani} and references therein).

On the other hand, for OBC most investigations deal with simple
one-component systems, especially the asymmetric simple exclusion process 
(ASEP) \cite{asep1,asep2,asep3,asep4,asep5}. In \cite{PSSS}
the NaSch model with $\vm >1 $ was studied with OBC and the results
compared to empirical data. Special boundary conditions for this
case were also studied in \cite{CKS} \footnote{See, however, 
the discussion of these results in \cite{newVDR}.}.
For driven lattice systems which exhibit  a metastable or bistable regime
for periodic boundary conditions not much is known about possible
phase diagrams in the case of open boundaries (see, however,
\cite{newVDR,AppSan}), nor have systems been studied with stable high 
flow branches.
%%
%%%%%%%%%%%%%%%%%%%%%%%%%%%%%%%%%%%%%%%%%%%%%%%%%%%%%%%%%%%%%%%%%%%%%
%%
\subsection{ASEP with open boundary conditions}
\label{ss_skob_asep}
The asymmetric simple exclusion process (ASEP) is the simplest
prototype-model of interacting systems driven far from equilibrium. It
is a generic model for studying driven systems and boundary-induced
phase transitions \cite{krug,sz,gs}.

The ASEP is a discrete particle hopping model. A particle can move
forward one cell with probability $p$ if the lattice site immediately
in front of it is empty.  If the first cell, corresponding to the left
boundary\footnote{We assume that the particles move from left to
  right.}, is empty a particle will be injected there with probability
$\alpha$. If the last cell is occupied the particle will be removed
with probability $\beta$. By varying the tuning parameters $\alpha$
and  $\beta$, and therefore the densities at the boundaries, one
obtains a surprisingly rich phase diagram.

One distinguishes three different phases according to the functional
dependence of the current and the corresponding stationary bulk density on
the system parameters. In the 
\emph{low-density phase} the current is independent of $\beta$. Here
the current is limited by the input rate $\alpha$ which then dominates
the behaviour of the system. In the \emph{high-density phase} the
behaviour is dominated by the output rate $\beta$ and the current is
independent of $\alpha$. In the \emph{maximum current phase} the
limiting factor for the current is the bulk rate $p$ and the current
becomes independent of both $\alpha$ and $\beta$. 

In \cite{Kolo98} a nice physical picture has been developed which
explains the structure of the phase diagram not only qualitatively,
but also quantitatively. By considering the collective velocity $v_c =
q'(\rho)$ which is the velocity of the center of mass of a local
perturbation in a homogeneous, stationary background of density $\rho$
and the shock velocity $v_s = \frac{q_2 - q_1}{\rho_2 - \rho_1}$ of a
`domain wall' between two stationary regions of densities $\rho_{1}$
and $\rho_{2}$, one can understand the phase diagram of systems with
unique flow-density relation from the fundamental diagram of the
periodic system \cite{Kolo98,popkov}. The idea behind is that these
two velocities determine if and how a perturbation will spread through
the system. For a detailed discussion see \cite{popkov}.

A general valid ``rule'' is found for systems with unique
flow-density relations, i.e. the current always obeys an {\em extremal
 current principle} \cite{Kolo98,popkov}:
\begin{eqnarray}   \label{expr}
  \begin{array}{lcl}
    q & = &
    \max_{\rho \in [\rho^{+},\rho^{-}]} q(\rho) \quad 
    \mbox{ for } \rho^{-}>\rho^{+}, \\[2ex]
    q & = &
    \min_{\rho \in [\rho^{-},\rho^{+}]} q(\rho) \quad
    \mbox{ for } \rho^{-}<\rho^{+}.
    \end{array}
\end{eqnarray}
$\rho^-$ and $\rho^+$ are effective densities at the left and right 
boundary, respectively. The principle (\ref{expr}) states
that the phase diagram of the open system is completely determined by
the fundamental diagram $q(\rho)$ of the periodic
counterpart. Moreover, it implies that two models with different
microscopic dynamics, but the same fundamental diagram, will have the
same phase diagram for open boundaries. In this sense the phase
diagram is independent of the microscopic dynamics. 
%%
%%%%%%%%%%%%%%%%%%%%%%%%%%%%%%%%%%%%%%%%%%%%%%%%%%%%%%%%%%%%%%%%%%%%%
%%%%%%%%%%%%%%%%%%%%%%%%%%%%%%%%%%%%%%%%%%%%%%%%%%%%%%%%%%%%%%%%%%%%%
%%%%%%%%%%%%%%%%%%%%%%%%%%%%%%%%%%%%%%%%%%%%%%%%%%%%%%%%%%%%%%%%%%%%%
%%
\subsection{Krauss-model with open boundary conditions}
\label{sec_skob}
The aim of this paper is the study of the Krauss-model with open
boundary conditions, especially obtaining its phase diagram, and furthermore 
to investigate its connection to the theory of boundary-induced phase
transitions (see
section~\ref{ss_skob_asep}). As will be shown, the choice of
appropriate injection/extraction strategies at the boundaries of the
system plays a crucial role. Here it should be kept in mind that the 
principle (\ref{expr}) is formulated in terms of effective boundary
densities which result from these strategies. 

The rules specified in the following sections have to be such that the
full range of possible bulk states (compare Fig.~\ref{fig:FDG_p}) may
be reached (at least theoretically). Especially, we are interested in
states of high flow. Therefore, one might think of a strategy to
inject cars with a initial velocity of $\vm$ and injection rate
$\alpha$. However, this will not lead to a crash-free motion by itself.

Since the model is known to be crash--free from the closed system,
this is somehow surprising. The reason can be found from the fact that
under open boundary conditions all kind of initial situations can
occur due to the stochastic feeding of cars. For the safety of the
model quantity $\xi(t) = g(t) - v_l(t)$ plays a
cruicial role. Its evolution for the deterministic Krauss-model ($\epsilon
= 0$) is given by \cite{skthes}
\begin{eqnarray}                                
  \label{xidev}
  \xi(t + \Delta t) \ge \xi(t) \left(1 - \frac{1}{\tau_b + 1} \right),
\end{eqnarray}
with $\tau_b = (v + v_l) / 2b$. Equation (\ref{xidev}) implies that if
once $\xi \ge 0$ (and as a result $g \ge v_l\ge 0$) this will hold for all
future timesteps. Safety is therefore guaranteed if $\xi(t = 0) \ge
0$. The latter condition is not fulfilled automatically if cars are
fed with a rate $\alpha$. In simulations  we found that a car
which collides with its predecessor always had $\xi (t=0) < 0$. Note
that the opposite is not true, i.e.\ a car that started with negative
$\xi$ does not have to be involved in a crash
\footnote{From (\ref{xidev}) follows that a crash is most probable for
  slow moving leaders and small initial gap $\gi$.}. Due to the stochastic 
step in the update rules (\ref{eq:THE model}), $\xi$ can be pushed from
negative to positive values but not the other way around.
Just choosing smaller initial velocities reduces the probability that
$\xi (t=0) < 0$ leads to a crash (even to negligible values) but the
states of high flow will not be reached \footnote{That is another
  difference to the NaSch-like models in which the choice of $v \le g$
  always leads to collision-free motion.}.

We follow two different strategies of injecting cars to overcome the
mentioned problems. One strategy is based on the following idea: If
there is enough space at the beginning of the system, i.e.\ at least one
carlength $\lc$, cars are injected according to the injection rate $\alpha$.
As a consequence one has to define an initial velocity $\vi = v_f (t =0)$ 
that is as high as possible (since in the high flow states the average 
velocity $v \approx \vm$), but keeps the system free of crashes.
The problem here is, when using the formula for $\vs$ from
(\ref{eq:THE model}), that a velocity $v_f$ is already needed for its
calculation. Moreover one has to deal with the cases in which $\xi
(t=0) \le 0$. Using the safety condition (\ref{eq:safety_gipps})
we will derive a rule to determine $\vi$ in section~\ref{sec_skob_alpha}.

The other strategy investigated goes the opposite way. The
high-current states of the Krauss-model are characterized by velocities
close to $\vm$ and more or less identical gaps $g \gg \lc$.  This
property can be used to define a rule which mimics the structure of
the high current states, i.e.\ one defines a minimal gap $\gi > \lc$
that has to be respected at the left boundary and injects all cars
with $\vi = \vm$. It should be noted that in this case the injection
rate does not equal $\alpha$ anymore, but becomes a monotoneously
increasing function of $\alpha$. Details and simulation results are
given in section~\ref{sec_skob_v}.
%%
%%%%%%%%%%%%%%%%%%%%%%%%%%%%%%%%%%%%%%%%%%%%%%%%%%%%%%%%%%%%%%%%%%%%%
%%%%%%%%%%%%%%%%%%%%%%%%%%%%%%%%%%%%%%%%%%%%%%%%%%%%%%%%%%%%%%%%%%%%%
%%%%%%%%%%%%%%%%%%%%%%%%%%%%%%%%%%%%%%%%%%%%%%%%%%%%%%%%%%%%%%%%%%%%%
%%
\section{An inflow-oriented injection rule}
\label{sec_skob_alpha}
In this section an injection method is introduced which is similar to
that for the ASEP. Cars are injected into the system with inflow rate
$\alpha$ whenever there is at least one carlength space in front of
the system ($\gi \ge \lc$), using any safe initial
velocity $\vi$ (depending on the system configuration). However, in
order to reach large currents, cars have to be injected with the
maximum safe velocity possible.

The rule, as stated up to now, leads the condition $\xi(t=0) =
\xi_{\rm init} < 0$ which can cause accidents as seen in
section~\ref{sec_skob}. Since we want to investigate a naive
generalisation of the injection strategy of the ASEP and to compare the 
results to it, it is necessary to think about a rule which is mainly
oriented on $\alpha$. Moreover, in real-world applications one has to
understand the behaviour of such a rule, since cars usually are
inserted according to a given inflow 
%at a link 
instead of particular strategies.
\subsection{Boundary rules ({\it rule~1})}
\label{sec_rule1}
In order to complete the rule we have to find the maximum safe
velocity $\vi$ possible. We can not just use (\ref{eq:safety_gipps})
since $v_f$ is not known. Moreover, the velocity has to be such that
the dynamics of the system allows the transition from $\xi_{\rm init} < 
0$ to $\xi \ge 0$. As a solution we do not look only at the first car 
in the system, but also at its predecessor.

The open boundary conditions for the inflow--oriented rule are defined 
in the following way:\\[0.1cm]
\noindent {\it Step~1: Injection}\\
If there is at least one car length free space at the beginning of the 
system, with probability ${\alpha}$ we inject a car with velocity
$\vi$: 
\begin{equation}
  \label{eq:Vi3}
  \vi = \min\left\{\vm,\sqrt{2bg+\frac{b}{b_l} v_l^2 + b^2} - b\right\} .
\end{equation}      
This velocity is a function of $\vm$, the velocity $v_l$ of
the leading car, the deceleration rate $b$ and an upper bound $b_l$
for the actual deceleration of the leading car. The latter is
calculated using the velocity of the car in front of the leading car
and therefore, gives a bound for the worst case, i.e., the maximum
deceleration of the leader in the next timestep.\\[0.1cm]
\noindent{\it Step~2: First update}\\
Performing the first update of a car injected at the current timestep,
we define an own rule. Given $\vi$ of {\it step~1} we follow the
update rules of the Krauss-model in case of the leading car moving with $v_l
> \vc$, where $\vc$ is a constant velocity depending on $\vm$. If $v_l
\le \vc$, $\vs$ is set equal to the gap $g$ instead of using
(\ref{eq:THE model}). This defines a cutoff for which $\xi_{\rm init} < 0$
still leads 
to safe driving while keeping $\gi$ close to $1$ for high values of
$\alpha$.\\[0.1cm]
\noindent {\it Step~3: Extraction}\\
With probability $1-{\beta}$ a block is added at the end of the road 
which causes the car at the end of the system to slow down.
Otherwise, with probability $\beta$, the cars simply move out of the
system.\\[0.1cm]
\noindent {\it Step~4: Update}\\
Update with the Krauss-model update rules (see section~\ref{sec_SK_Model}).\\

In the following we use parameters as given in (\ref{eq:params}) and
$\vc=1.6$, $\gi=\lc=1.0$. The value for $\vc$ has been determind by
means of simulation. Note that $\gi$ is the space that has to be
free at least at the left border of the system.
%%
%%%%%%%%%%%%%%%%%%%%%%%%%%%%%%%%%%%%%%%%%%%%%%%%%%%%%%%%%%%%%%%%%%%%%
%%
\subsection{Fundamental diagram}
\label{ss_skob_alpha_fd}
In contrast to periodic boundary conditions, the fundamental diagram
$q(\rho)$ is not easy to find for the full range of bulk densities.
While for a closed system the density $\rho$ is given and conserved, in
open system it is a quantity that results from the parameters $\alpha$
and $\beta$. Their influence on $\rho$ or $q$ is non--linear. Another
difficulty is that global density and current should be measured in
the stationary state which is reached after quite long simulation
times for certain values of $(\alpha,\beta)$. A detailed discussion
can be found in \cite{NamaziDIP}. Because of the complex relation
between $(\alpha,\beta)$ pairs and $\rho$ or $q$, one can not find a
value for the $q$ for each $\rho$ and vice versa. It should be
mentioned that $\rho$ and $q$ are rather sensitive to changes in
$\alpha$ or $\beta$ \cite{NamaziDIP}.

Our strategy in finding the fundamental diagram is as follows: For
each pair of parameters $(\alpha, \beta)$, the simulation has to be
run until the stationary state is reached. Then, the flow $q$ and bulk
density $\rho$ are measured in the middle of the system. For the
latter this has been done in an interval of length $\frac{L}{3}$.
These two values fix a point in the $\rho$--$q$ plane. To make the
data more reliable and to reduce the influence of noise, this has been
repeated for several times with different random seeds.

\begin{figure}[t]
\centerline{\epsfig{figure=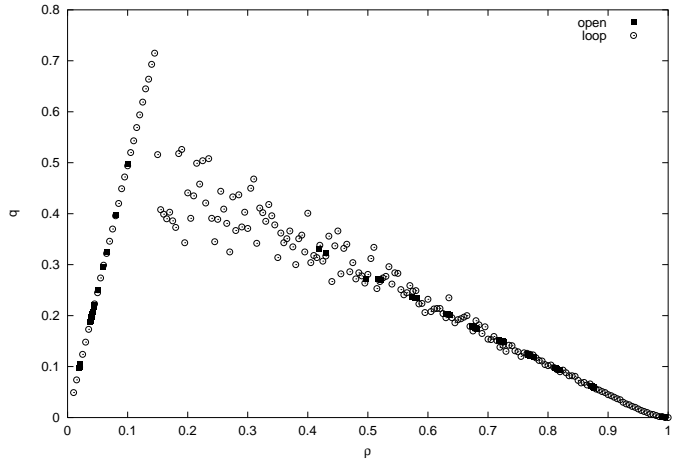,width=9cm}}
\caption{\protect{Comparison between the fundamental diagram of the
    open system ($\blacksquare$) and the one from the closed system
    ($\circ$). The points are obtained from density profiles in
    case of the open system.}}
  \label{fig:fdg_c}
\end{figure}

In Fig.~\ref{fig:fdg_c} the fundamental diagram of a system with
periodic boundary conditions and the same system with open boundary
conditions are compared. The high-flow branch in the bistable density
regime does not exist in the open case. There are also other states 
missing in the fundamental diagram
of the system with open boundary conditions, i.e., certain densities
can not be generated by the boundary rules defined in
section~\ref{sec_rule1}.
%%
%%
%%
%%%%%%%%%%%%%%%%%%%%%%%%%%%%%%%%%%%%%%%%%%%%%%%%%%%%%%%%%%%%%%%%%%%%%%%%%%%%%%%
%%
\subsection{Phase diagram}
\label{ss_skob_alpha_ph}
The phase diagram as a function of $\alpha$ and $\beta$ can be
obtained by studying the density profiles of systems in the stationary
state. It is easy to distinguish between low- and high-density
phases. To find the maximum current phase one should compare the
($\rho$, $q$) pair of the system with the fundamental diagram of the
same system with periodic boundary conditions.

\begin{figure}[t]
  \centerline{\hspace{3cm}\epsfig{figure=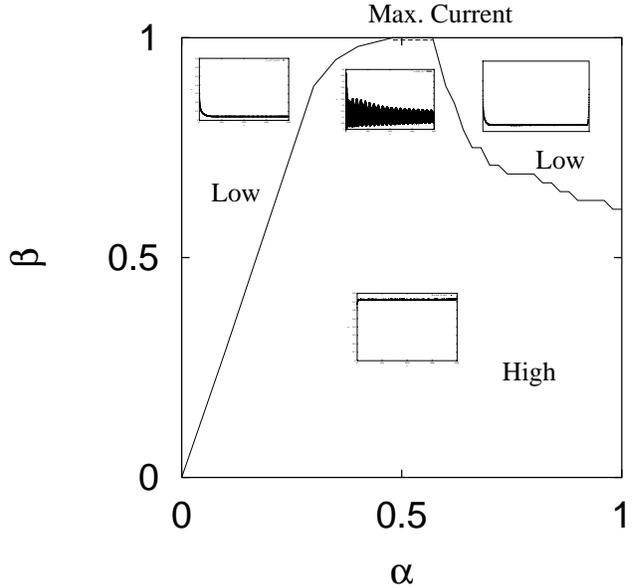,width=12cm}}
  \caption{\protect{Phase diagram for the Krauss-model with open boundary
  conditions and the inflow-oriented injection strategy. The inserts
  show typical density profiles for each phase.  The maximum current
  phase with flow $q \approx 0.5$ is only observed for systems with
  open right boundary ($\beta=1$) and injection rate between
  $\alpha=0.48$ and $\alpha=0.57$ (broken line) and shows an oscillating 
  density profile.}}  \label{fig:PD}
\end{figure}

Using the density profiles for systems with $\alpha,\beta \in [0,1]$
with increments of $0.02$, we have drawn a phase diagram in
Fig.~\ref{fig:PD}. Different phases, phase boundaries and the typical
density profiles for each phase are shown. The maximum current phase
with $q$ about $0.5$ has been observed for system with open right
boundary ($\beta=1$) and injection rate between $\alpha=0.48$ and
$\alpha=0.57$ (broken line). Note that all stable high-flow states
in the bistable region correspond to some point in the maximal current
phase.

This phase diagram looks different from phase diagrams of similar
models since one finds two different low density regimes. 
Therefore, by varying $\alpha$ with $\beta$ kept constant, one
observes a reentrance transition for large values of $\beta$.
It is natural to assume that this is related to the special choice 
of input and output strategies. To verify this, later on
(Sec.~\ref{sec_skob_v}) a different injection strategy will be
studied. The reason why this phase diagram looks unusual can be
understood studying the relations between $\alpha$, $\beta$ and
dynamic parameters of the system (see Sec.~\ref{ss_skob_alpha_aq}).

A similar phenomenon has been observed in \cite{antal} in a simple
one-component lattice gas with next-nearest neighbour interaction.
Here a reentrance transition to a second high-density phase was
found. Although the origin of this transition is not entirely clear,
the authors of \cite{antal} argued that it is related to the complicated
connection between boundary rates and the effective boundary densities.
%%
%%%%%%%%%%%%%%%%%%%%%%%%%%%%%%%%%%%%%%%%%%%%%%%%%%%%%%%%%%%%%%%%%%%%%
%%
\subsection{Dependence of dynamic quantities on ${\alpha}$,
  ${\beta}$}
\label{ss_skob_alpha_aq}
\begin{figure}[t]
  \centerline{\psfig{figure=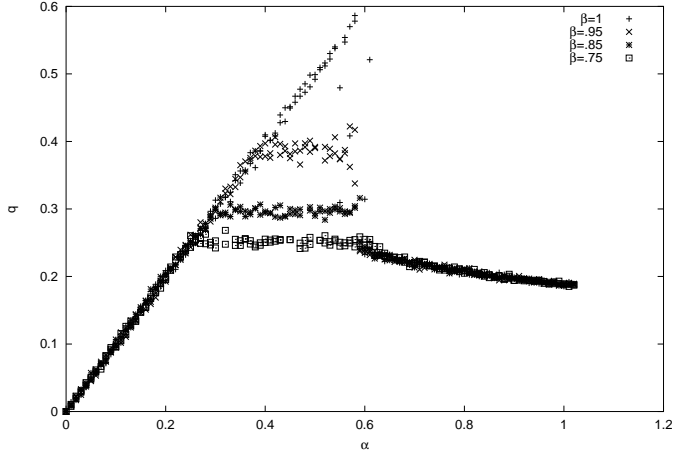,width=9cm}}
  \caption{\protect{$\alpha-q$-diagram for $\beta \in [0.70,1]$. The
  maximal reached current decreases for smaller $\beta$ since jams
  are builded due to the disturbance at at the right boundary of the
  system. The strong drop of the current for $\alpha > 0.58$ is a
  result of the cutoff $\vc$ in {\it step~2}.
  }}
\label{fig:aqs}
\end{figure}

The dynamics of a system with open boundary conditions depends on the
parameters $\alpha$ and $\beta$. When $\beta=1$, i.e.\ the outflow is
unrestricted, an increasing flow can be expected with increasing
injection rate ${\alpha}$. In Fig.~\ref{fig:aqs} the flow of the
stationary state is plotted versus $\alpha$. However, this is true
only for $\alpha < 0.58$. For larger values, a sharp decline in the
flow can be seen which can be related to the injection strategy.
Note that the second injection strategy introduced in
section~\ref{sec_skob_v} does not show that decline in the
$\alpha-q$--relation. 

For systems with $\beta \neq 1$ the right boundary introduces an 
external disturbance which increases with decreasing
$\beta$ (cf.\ {\it step~3}). With ${\beta} \neq 1$, the cars at the
end of the system are forced to slow down. Hence, the density at the
right boundary increases with decreasing $\beta$ and jams are
formed. These jams grow backwards into the system. If the jams can not
dissolve due to high inflow rates $\alpha$, the cars have a lower
average velocity and the system's density is high. 

Because of the bistability, the system has quite a different behaviour
for $\beta = 1$ in comparison to any other $\beta$. Note that already
small (external) perturbations might force the breakdown of the
stable high-flow states in the open system. Once a jam has
established in the system 
the high-flow states will not be reached again due to the reduced
outflow from jams, $(\rho,q_{\rm out}) = (0.11,0.51)$, which is smaller
than the maximal possible flow. As an example for the behaviour of the
system under a weak disturbance, the $\alpha-q$-relationship 
for $\beta = 0.95$ is shown in Fig.~\ref{fig:aqs}.

On the other hand, one can study the relationship between $q$ and
$\beta$, using a constant value of $\alpha$ (not shown). It is
obvious that for very small values of $\beta$, the current should
vanish. In a system with closed right boundary ($\beta=0$), all cars
are forced to stay in the system which means a vanishing average
velocity $v = 0$ and density $\rho=1$. For each $\alpha$, one
expects the highest value of current for $\beta=1$.

\begin{figure}[t]
  \centerline{\epsfig{figure=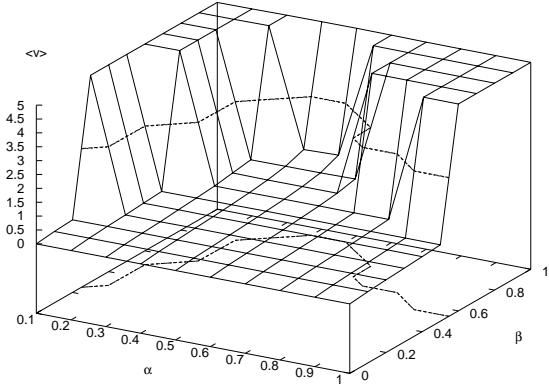,width=9cm}}
  \caption{\protect{Average velocity as a function of $\alpha$ and
  $\beta$. The contour line (-~-~-), given by $v > \vm /2$, separates
  the free flow phase (average $v$ close to $\vm$) from the
  jammed phase ($v$ much smaller then $\vm /2$). 
  }} \label{fig:v3d}
\end{figure}

In Fig.~\ref{fig:v3d} the average velocity as a function of $\alpha$
and $\beta$ is shown. The contour in the $\alpha$--$\beta$--plane
shows the line where $v = \vm / 2$. In \cite{skthes} this line was
choosen to distinguish between jam 
and free flow, i.e. states with $v < \vm / 2$ are in the jammed
phase. 

\begin{figure}[t]
  \centerline{\psfig{figure=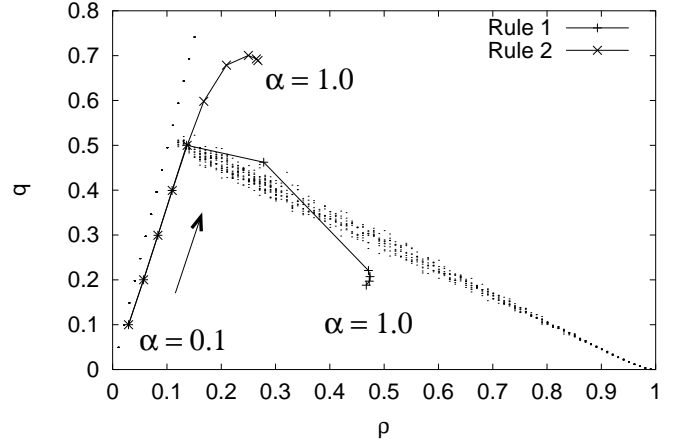,width=9cm}}
  \caption{\protect{Simulation of the left boundary rules
      given in section~\ref{sec_rule1} (Rule~1) and~\ref{sec_rule2}
      (Rule~2). In each timestep only the leftmost three cars are
      simulated while the rest of the system is cutted off. The dots
      represent the fundamental diagram taken from the closed system.
      After a relaxation of $100000$ timesteps the points $(\rho,q)$
      are measured over another $100000$ steps for different values of 
      $\alpha$. The arrow indicates the direction of increasing $\alpha$.
      }}
  \label{fig:lb}
\end{figure}

Before we will examine the connection to the extremal principle the
effects of {\it rule~1} are investigated in more detail. In order to
measure the impact of that rule on flow and density we let the system
run for different values of $\alpha$. In each timestep only three cars
are left in the system by taking out the rest without taking care on
their position or velocity, i.e., the rest of the system is cutted
off. The resulting relation between the density and flow at the left
boundary is shown in Fig.~\ref{fig:lb}. Following the line starting
from $\alpha = 0.1$ the left border stays on the free flow branch of
the fundamental diagram up to $\alpha = 0.6$. For bigger values the
system switches to just one state that belongs to the jammed
branch. The drop in the flow is therefore not
alone a result of jams moving backwards to the left boundary, but an
artificial effect of the rule itself. If the predecessing vehicles
are moving slow ($v < \vc$)
\footnote{which results e.g. from jam formation around the left boundary}
the safe velocity of the inserted cars becomes $g$ due to {\it step~2}
of {\it rule~1}. And, $g$ will be very small in case of high injection 
rates. These effect will play a role for the interpretation of the
phase diagram in the context of the extremal principle.
%%
%%%%%%%%%%%%%%%%%%%%%%%%%%%%%%%%%%%%%%%%%%%%%%%%%%%%%%%%%%%%%%%%%%%%%
%%
\subsection{Extremal principle}
\label{ss_skob_alpha_EP}
\begin{figure}[t]
  \centerline{\psfig{figure=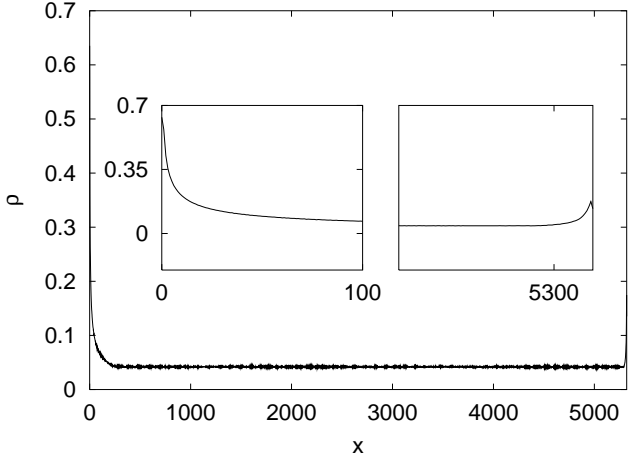,width=9cm}}
  \caption{\protect{Density profile for the system with
  $(\alpha,\beta) =(0.8,0.8)$ which lies in the second low density
  phase of Fig.~\ref{fig:PD}. The inserts show a close up at the
  boundaries. The densities at the left and right boundary are $\rho^-
  = 0.65$ and $\rho^+ = 0.17$ respectively, the bulk density is
  $\rho_{\rm b} = 0.04$. A stationary flow $q = 0.21$ is measured.  }}
  \label{fig:profile}
\end{figure}

Since we could not observe the full range of states
(cf.\ Fig.~\ref{fig:fdg_c}) due to the cutoff in {\it rule~1} at a
certain inflow we examined the extremal principle (see
Sec.~\ref{ss_skob_asep}) only for a subset of system states. Since the
extremal principle (\ref{expr}) is formulated in terms of boundary
densities these have to be determinded from the stationary density
profiles. For small injection rates it is not easy at all to find
the left boundary density $\rho^-$, due to the oscillations in density
profiles. This is no problem for models that have $\vm = \lc$ 
(= particle size).

For the parts of the phase diagram that look similar to other models,
i.e., all except for the second low density regime $\alpha > 0.6$,
$\beta > 0.6$ (cf. Fig.~\ref{fig:PD}), it should just be stated that
the principle was fulfilled for all these pairs of $(\alpha,\beta)$.

A different behaviour is found for the second low density regime.  The
results of section~\ref{ss_skob_alpha_aq} suggest that this is due to
the fact that for {\it rule~1} we do not have a monotonic increase of
inflow with $\alpha$. Consider the point $(\alpha,\beta) =(0.8,0.8)$
which lies in the second low density phase. Fig.~\ref{fig:profile}
shows the corresponding stationary density profile.
Using the formulation (\ref{expr}) one finds an apparent violation of
the extremal principle, as will be demonstrated. From the density
profiles we find $\rho^- = 0.65$ and $\rho^+ = 0.17$ and since $\rho^-
> \rho^+$ the maximum of $q(\rho^-)$ and $q(\rho^+)$ will be chosen
according to (\ref{expr}), i.e., $q^{(\text{pred})} = 0.48$. Instead
we measure a stationary flow of $q= 0.21$.

Before we present an alternative interpretation of the results in the
second low density phase we recall some findings in the context of the 
ASEP. All the results have been obtained for models with $\vm = 1 =
\Delta t$. Therefore, in these models the flow obtruded to the system 
by the boundary rules does play no role. A stopped particle can
accelerate to $\vm=1$ in one timestep. Then, the
formulation as given in (\ref{expr}) is sufficient to determine the
system state. For models with $\vm > 1$ (cf. \cite{newVDR}) the left
boundary rule has to be defined in a specific way to obtain results
compatible with the extremal principle. The rule used in \cite{newVDR}
always allows the injection of cars with $\vm$. Hence large flows
can be reached and the problem that acceleration to the maximum 
velocity takes several timesteps does not occur.

In our case the left boundary density is large due to the high
injection rate ($\rho^- \propto \alpha$) but the flow
is restricted to $q = 0.21$ due to the cutoff in {\it step~2} of {\it
rule~1} (cf. Fig.~\ref{fig:lb}). Therefore, the system has to choose a
state that matches with that flow. Since the exit allows a higher flow
no stable growing jam can develope in the system and the state on the
free flow branch is choosen. Indeed the profile shows a sharp decrease
in the density leading to a bulk density $\rho_{\rm b} = 0.04$ for
which $q(\rho_{\rm b}) = 0.21$ (cf. Fig.~\ref{fig:FDG_p}). 
%Hence we claim that in the case of time--discrete models for which $\vm$ can
%not be reached in one timestep the principle (\ref{expr}) can not be
%formulated only taking into account the boundary densities $\rho^-$, $\rho^+$.
This indicates that the extremal principle (\ref{expr}) formulated
only in terms of boundary densities $\rho^-$, $\rho^+$ is not sufficient.
Moreover, one has to check if there is a restriction due to
the inflow and outflow rules (denoted by $q^{-/+}$). (\ref{expr}) then only
applies for $q \le q^{-/+}$. Otherwise the system state is chosen by
\begin{eqnarray}                                
  \label{jp}
  q_{b} & = & \min \{q^-(\alpha),q^+(\beta)\}
\end{eqnarray}
with a density $\rho_{b}$ satisfying $q(\rho_{b}) = q_{b}$. Predicting
the stationary system state according to (\ref{expr}) together with
(\ref{jp}) one finds good agreement for all pairs of
$(\alpha,\beta)$. The results given for another injection strategy as
formulated in section~\ref{sec_skob_v} will confirm this
interpretation. We believe that similar extensions of the extremal
principle will be necessary for other multicomponent models.
%%
%%
%%%%%%%%%%%%%%%%%%%%%%%%%%%%%%%%%%%%%%%%%%%%%%%%%%%%%%%%%%%%%%%%%%%%%
%%%%%%%%%%%%%%%%%%%%%%%%%%%%%%%%%%%%%%%%%%%%%%%%%%%%%%%%%%%%%%%%%%%%%

\section{High-velocity-oriented injection rule}
\label{sec_skob_v}

As seen in the last section it is not easy to define left boundary
conditions that allow the system to reach states belonging to the
stable high-flow branch and -- at the same time -- ensure safe
motion of the injected car under any circumstances. In the following a
rule is formulated that tries to inject cars with the maximal
velocity $\vm$ of the model.
\subsection{Boundary rules ({\it rule~2})}
\label{sec_rule2}
In order to generate the high-flow states we have a closer look at the 
bistable regime of the corresponding periodic system. Here high-flow states
are characterized by velocities close to $\vm$ and approximately identical
gaps $g \gg \lc$. Therefore, instead of driving with initial speed
$\vi$ to achieve safety, whenever there is an empty space of one carlength
$\lc$ in front of the system, we try to inject cars 
%\footnote{Depending moreover on the inflow rate $\alpha$.}
(with probability $\alpha$) with an initial velocity $\vm$. 
To guarantee a system
free of crashes, we have to introduce a minimum safety gap $\gi$ to
the preceding car (cf.\ also the discussion in section~\ref{sec_skob}).
Therefore not all injection trials will be successful (see below).

{\it Step~1} and {\it step~2} of {\it rule~1} are replaced by above strategy
while the definition of the right boundary conditions is not
changed. Using this strategy, it might happen that no car will be injected 
for several timesteps because of the lack of free space ($\gi$) at the 
beginning of the system. This implies that the actual injection rate
is smaller than $\alpha$ and also the existence of a maximal 
density $\frac{1}{\gi +1}$ which can be reached at the left boundary.

For the choice of the initial gap $\gi$ three criteria are formulated:
\begin{description}
\item[$(i)$] $\gi$ should leave the system crash-free.
\item[$(ii)$] High currents (comparable to those found in
the bistable regime of the periodic system) should be reached.
\item[$(iii)$] $\gi$ should be as small as possible.
\end{description}

The last criterion increases the maximum reachable density at
the system's entry so that a larger range of boundary densities can
be investigated.

To find an appropriate value for the initial gap, we made several
simulations with different values for $\gi$, and measured the current
$q$ for $\beta=1$ to test criterion $(ii)$. 
For the values of $\gi$ which satisfy $(ii)$ we then checked in
simulations whether the system is free of crashes.
%Then we made some simulations with this set and tried to
%observe crashes in the system to filter out unsuitable values. 
This could be done best using $\beta \approx 0$. After filtering out
unsuitable values of $\gi$ we chose the minimum of the remaining values.

\begin{figure}[t]
\centerline{\epsfig{figure=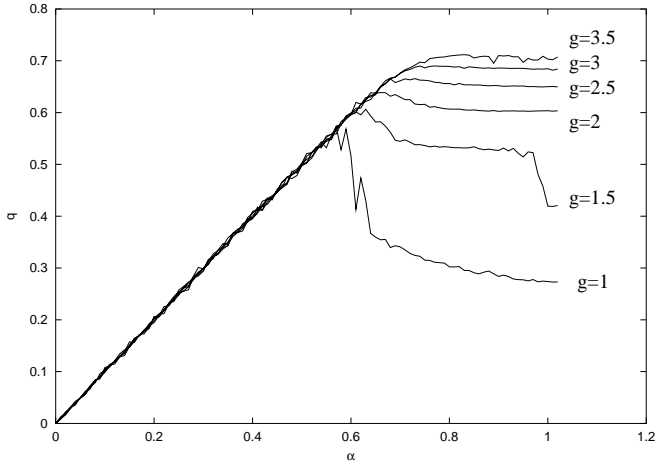,width=9cm}}
\caption{\protect{The relation between the current $q$ and $\alpha$
    for several values $\gi \in [1,3.5]$, $\beta = 1$ and
    the parameters equal to (\ref{eq:params}).
    }}
\label{fig:qa_gap2d}
\end{figure}

In Fig.~\ref{fig:qa_gap2d} the relation between $q$ and $\alpha$ for 
several values of $\gi$ and $\beta=1$ is shown. The current
increases as $\alpha$ increases and either reaches a constant value or
decreases and then reaches a constant value, depending on $\gi$. For
$5 > \gi \ge 2 $ the current reaches a constant value greater than
$0.6$, which is in the bistable region of a system with periodic
boundary conditions.  For $\gi \ge 5$ $q$ decreases drastically. This
is obvious, since it will be impossible to reach a density in the
regime of maximum current in that case. The maximum current in terms
of the periodic system will be reached for a system with $\beta=1$ and
density of $0.15$.

In simulations for systems with $\beta=1$ and $\beta\ll 1$ we have
observed crashes for $\gi < 2$. Therefore, the only values of $\gi$
meeting the criteria $(i)$, $(ii)$ are values $5 > \gi \ge 2$. Because
of the criterion $(iii)$ we decided to take $\gi = 2$ in order to
reach a maximum possible left boundary density.
 
In the following we present results of simulations using this
injection rule, parameters as given in (\ref{eq:params}) and $\gi =
2, \vi = \vm$. As long as the methods and interpretations correspond
to section~\ref{sec_skob_alpha} the presentation will be kept brief.
%%
%%%%%%%%%%%%%%%%%%%%%%%%%%%%%%%%%%%%%%%%%%%%%%%%%%%%%%%%%%%%%%%%%%%%%
%%
\subsection{Fundamental diagram}
\label{sss_fdg_g2}
\begin{figure}[t]
\centerline{\epsfig{figure=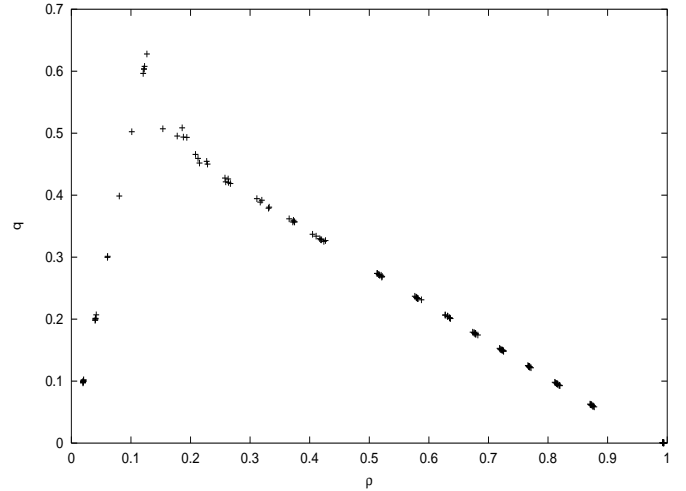,width=9cm}}
\caption{\protect{Fundamental diagram of the system with open
    boundaries and boundary conditions defined by {\it rule~2}.  The
    maximum current and the high-flow states in the bistable region
    are reached for     systems with $\beta=1$. The fundamental diagram 
    is similar to the fundamental diagram of the system with periodic 
    boundary conditions (see section~\ref{sec_SK_Model}).}}
\label{fig:fdg_gap2}
\end{figure}

The fundamental diagram obtained for this system is very similar to
the system with periodic boundary conditions. The high-flow
states in the bistable regime can be reached using $rule~2$, see
Fig.~\ref{fig:fdg_gap2}. With the present injection strategy,
increasing $\alpha$ does not cause the decrease of the current in the
way we saw in section~\ref{sec_skob_alpha}. The high-flow states are
only reached for systems with $\beta=1$ (cf.\ next subsections).
%%
%%%%%%%%%%%%%%%%%%%%%%%%%%%%%%%%%%%%%%%%%%%%%%%%%%%%%%%%%%%%%%%%%%%%%
%%
\subsection{Dependence of dynamic quantities on ${\alpha}$,
  ${\beta}$}
\label{sss_qrv_g2}
In this section we give a brief overview of the dependence of the
current $q$, average velocity $v$ and the density $\rho$, measured in
the middle of the system, on $(\alpha,\beta)$. 

\begin{figure}[t]
  \centerline{\epsfig{figure=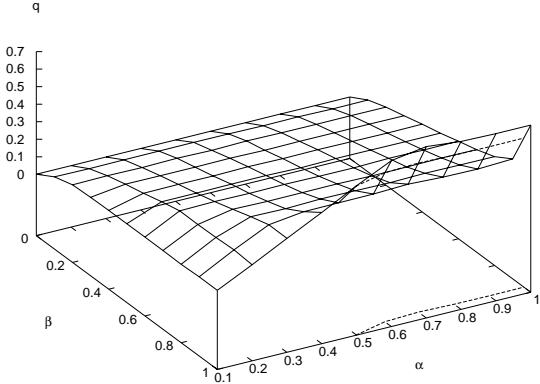,width=9cm}}
  \caption{\protect{The relation between the $q$, $\alpha$ and
  $\beta$ ($\gi=2$). There is no decrease of the current with
  increasing $\alpha$ anymore. The broken line correponds to $q = 0.5$.
  }}
\label{fig:qab_gap}
\end{figure}

In Fig.~\ref{fig:qab_gap} the relation between $q$, $\alpha$ and
$\beta$ is shown. The current increases strongly for $\beta$ values
close to $1$. We have also indicated the boundary between the
regions with $q>0.5$ and $q<0.5$.

\begin{figure}[t]
  \centerline{\epsfig{figure=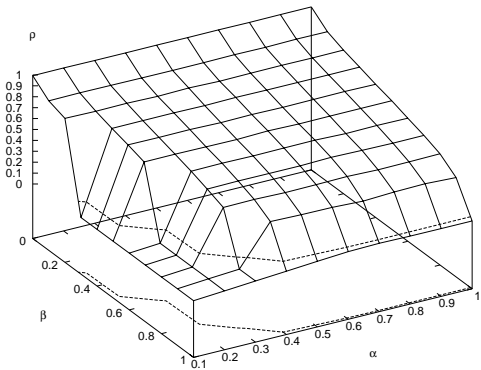,width=8cm}}
  \caption{\protect{The relation between the density $\rho$ (measured in
  the middle of the system in stationary state), $\alpha$ and $\beta$
  ($\gi=2$).  The broken line, corresponding to a density $\rho=0.15$,
  can be interpreted as a phase transition line between high and low 
  density phases.  
%{\bf (neu  machen!!! Eine Kontourline mit ihrer Definition!!!)}.
  }}
\label{fig:rab_gap}
\end{figure}

Fig.~\ref{fig:rab_gap} shows the dependence of the density on
$(\alpha,\beta$). The density used here is measured in the
middle of a system in the stationary state, which is the density
corresponding to the plateau in the density profile for most pairs
$(\alpha,\beta$). One cleary sees a sharp transition from low to high
values which will be interpreted as the phase transition line between
high and low density phases.

\begin{figure}[t]
  \centerline{\epsfig{figure=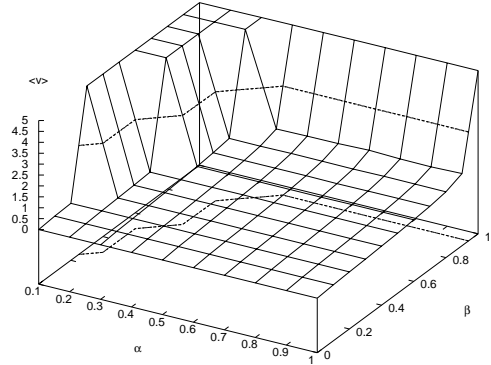,width=8cm}}
  \caption{\protect{The relation between the average velocity, $\alpha$ and 
      $\beta$ ($\gi=2$). The broken line corresponds to $v = \vm /2$ and 
      can be interpreted as transition line between free flow and jammed
      phase. }}
\label{fig:vab_gap2}
\end{figure}

Finally in Fig.~\ref{fig:vab_gap2} the dependence of the average
velocity on $(\alpha,\beta)$ is presented. Using again the criterion
$v = \vm / 2$ to distinguish between free flow and jam one finds the
same results for the phases as in Fig.~\ref{fig:rab_gap}.

%%
%%%%%%%%%%%%%%%%%%%%%%%%%%%%%%%%%%%%%%%%%%%%%%%%%%%%%%%%%%%%%%%%%%%%%
%%
\subsection{Phase diagram}
\label{ss_skob_v_ph}
\begin{figure}[t]
  \centerline{\hspace{3cm}\epsfig{figure=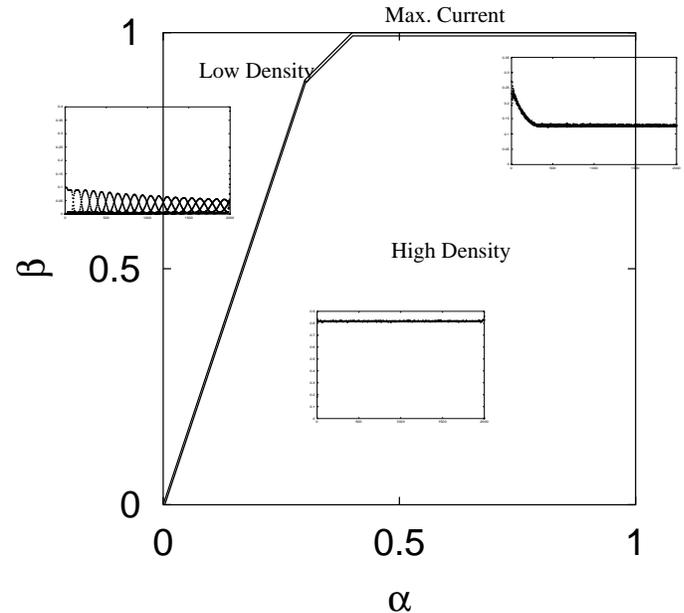,width=12cm}}
  \caption{\protect{Phase diagram of the Krauss-model with open
  boundary conditions and maximum-velocity-oriented injection rule
  ({\it rule~2}).  The maximum current phase has been observed for
  systems with open right boundary (${\beta}=1$) and $\alpha \ge 0.5$. 
  The inserts show (from left to right) typical density profiles for low, 
  high and maximum current phase. }}  \label{fig:PD_V}
\end{figure}

In Fig.~\ref{fig:PD_V} the phase diagram and different phases 
are shown. The diagram has been derived from the density
profiles. The full $\alpha$--$\beta$--plane was scanned in steps of
size $0.1$. The
maximum current phase has been observed for systems with open right
boundary (${\beta}=1$). It is reached for $\alpha \ge 0.5$. As one
can see from Fig.~\ref{fig:lb} at this value cars are fed in the
system with the flow out of jam, i.e., $q_{\rm in} (\alpha = 0.5) =
0.51 = q_{\rm out}$ \footnote{Using $\alpha = 0.5$ and $\beta
\in[0,1]$ one obtains the full high density branch of the fundamental
diagram.}.

This phase diagram is similar to that of the much simpler driven
lattice gas discussed in \cite{AppSan}. 
The results compare as well to the phase diagram found
for the VDR--model \cite{newVDR}. Therefore, we will only briefly
discuss the major differences found.

For the driven lattice gas of \cite{AppSan} the maximum current phase is only 
found for very short system. The reason is the true metastable nature of 
the high-flow states, i.e. intrinsic fluctuations are able to destroy
these states even without an {\em external} disturbance. Since the
probability for such fluctuations grow with the system length the
maximum current phase will vanish above a typical system size. In our
case, we do not find such a disappearance, even for very large
systems \footnote{We checked this for systems up to $L =
  50000$.}, since the high-flow states of the Krauss-model are stable
(as found in \cite{NKW}). 

The fact that we find the maximum current phase only for $\beta = 1$,
while in the case of the VDR-model it exists for a slightly bigger
range of $\beta$, can be related to the following. The interaction in
the VDR-model is very short-ranged due to the unbounded deceleration
capability in that model (the interaction horizon is $g \approx
\vm$). Instead, for the Krauss-model of the investigated class $I$,
cars do already interact for $g \approx \vm^2$.
%%
%%%%%%%%%%%%%%%%%%%%%%%%%%%%%%%%%%%%%%%%%%%%%%%%%%%%%%%%%%%%%%%%%%%%%
%%%%%%%%%%%%%%%%%%%%%%%%%%%%%%%%%%%%%%%%%%%%%%%%%%%%%%%%%%%%%%%%%%%%%
%%
\subsection{Extremal principle}
\label{ss_skob_2_EP}
We examined the validity of the extremal principle for this model for
every $\alpha$ and $\beta$ $\in[0,1]$ with $0.1$ steps. After
measuring the boundary densities $\rho^-$ and $\rho^+$ we have used
(\ref{expr}) to determine the current $q^{(\text{pred})}$ predicted by
the extremal principle. This value has then been compared to the
current $q$ measured in the open system. For all parameters the
results are in excellent agreement with $q^{(\text{pred})}$. This
is in accordance with our suggestion (see section~\ref{ss_skob_alpha_EP})
that as long as the boundary rules do not restrict the flow in an
artificial way, the system will choose its bulk state according to the
extremal principle as formulated in (\ref{expr}).
We also checked our formulation with the addition given in
(\ref{jp}). For all pairs of $(\alpha,\beta$) the correct bulk state was
predicted. 
%%
%%%%%%%%%%%%%%%%%%%%%%%%%%%%%%%%%%%%%%%%%%%%%%%%%%%%%%%%%%%%%%%%%%%%%
%%%%%%%%%%%%%%%%%%%%%%%%%%%%%%%%%%%%%%%%%%%%%%%%%%%%%%%%%%%%%%%%%%%%%
%%
\section{Conclusions}   
\label{SumCon}
In this article we investigated the properties of a traffic flow model 
introduced by Krauss et al. \cite{krauss,skthes,krauss2} under open
boundary conditions. It is a discrete map in time while --- in
contrast to cellular automata approaches --- space is continuous. 
For the parameter range discussed, the Krauss-model shows a
non-unique relation between flow and density. This property of 
the fundamental diagram is responsible for the stability of jams  
found in empirical observations. It is important to note that the
corresponding high-flow states have been shown to be stable \cite{NKW}
(subject to the model's dynamics). In contrast in cellular automata models
as the VDR-model \cite{VDR,newVDR} they usually show metastable
behaviour. 

For application purposes open boundary conditions play a cruicial role. 
But also for purely theoretical purposes the investigation of the 
Krauss-model's behaviour under open boundary conditions is valuable. 
In general, driven interacting particle systems show
{\em boundary-induced phase transitions}. For systems with unique 
flow-density relation, e.g.\ the ASEP as the prototype of such models, 
there exists a quite general theory for the stationary state the system 
realized with open boundary conditions (cf.\ section~\ref{sec_ob}). 
These results are well established for simple systems
with unique flow-density relation \cite{gs,Kolo98,popkov}. 
In contrast, not much is known about for systems with a non-unique 
fundamental diagram.
Moreover, the models investigated so far have $\vm = 1$ (i.e.\ $\vm$ 
can be reached within one timestep). Only recently also discrete models 
with higher velocities (e.g.\ models of Nagel-Schreckenberg type \cite{PSSS}) 
and non-unique flow-density relation \cite{newVDR,AppSan}
have been investigated. Here for the first time
a {\em continuous} model with $\vm \neq 1$ and fundamental diagram
with {\em bistability}, i.e.\ stable high-flow states,
has been studied with respect to boundary-induced phase
transitions.

In order to guarantee that the system remains free of crashes, one has 
to find a strategy to inject cars into the system. In this article two 
methods are discussed. The first one is orientated on the inflow rate
$\alpha$, i.e.\ with probability $\alpha$ a car is always inserted if there 
is enough space at the left boundary. In this case one has to find an
initial velocity which guarantees a crash-free motion of each inserted 
car  at any timestep (cf.\ section~\ref{sec_skob_alpha}). 
The second method fixes a minimum free space $\gi$ at the beginning of the 
system ($\gi > \lc$) and cars are injected with the constant velocity $\vm$
(cf.\ section~\ref{sec_skob_v}).

For both rules the phase diagram has been derived from computer
simulations and, as expected, boundary--induced phase transitions were
found. From the ASEP three different phases are known, distinguished
by the functional dependence of the current through the system on
$\alpha$ and $\beta$. These are the low--density, the high--density and 
the maximum--curent phase. All these phases are observed in the
Krauss-model with open boundary conditions. 

In contrast to the findings in the ASEP \cite{asep1,asep2,asep3,asep4,asep5}
the maximum current phase was only observed for an open right
boundary ($\beta=1$). This reflects the high sensitivity of the Krauss-model 
to an external disturbance at the system's exit which results from the long 
interaction horizon of the model. However, in contrast to \cite{AppSan}
the maximum current phase in the Krauss-model exists for arbitrary system 
lengths due to the stability of the high-flow states.

Moreover, the extremal principle (\ref{expr}) \cite{Kolo98,popkov} has 
been checked for both rules. For systems with $\vm = 1$ it was found
that the selection of the system's state does not primarily depend on
the parameters $(\alpha,\beta)$, but on the resulting densities
$\rho^-$ and $\rho^+$ at the left and right boundary,
respectively. Furthermore, the selected state is completely
determined by the flow-density relation of the corresponding periodic
system. In that sense the extremal principle states implies the
independence of the system state from the specific injection/extraction 
rule at the boundaries. For the second rule we found absolut agreement for the
Krauss-model (recall that here several timesteps are needed to accelerate
to $\vm$).

However, for the first rule deviations were found from the principle. 
An unusual reentrance transition to a second low-density phase occurs 
for high insertion and extraction rates. Its existence could be ascribed 
to the behaviour of the rule which shows a 
cutoff in the maximum flow for $\alpha \ge 0.6$. This flow is lower
than the one that is allowed by the right boundary conditions. In other 
words, for a model with $\vm \neq 1$ the selection of the system's
state is not independent of the specific boundary rules. We therefore
argue that the extremal principle as given in \cite{Kolo98,popkov}
can only be applied in this strict form to models with $\vm =1$. 
Here the provided maximum flow by the rule plays no role, since
a stopped car can always accelerate to $\vm$ in one timestep. 
Therefore any flow limitation due to the insertion rule does not
play a role.
This is different in the case of our model which has $\vm >1$ and finite 
acceleration capability. However, if one takes an additional rule into
account, i.e.\ that the flow in the system can not exceed the flow allowed
by the boundary rules, one can also predict the system state with open 
boundary conditions for our first rule (cf.\ section \ref{sec_skob_alpha})
from the knowledge about the periodic system.

We will close with a brief outlook. Due to our findings more work
on boundary--induced phase transitions for models with $\vm > 1$
seems to be necessary. The focus should be on finding a
compact formulation of the extremal principle that is independent of
$\vm$. Since in the Krauss-model crash-free motion is not included as a
rule per se (as, through hard-core exclusion, in the cellular automata 
approaches) it would be
helpful to develop a model formulation that does not show such safety
problems under open boundary conditions. This would be important for
applications in traffic flow simulation. To take into account
next-nearest-neighbor interactions as suggested by the findings of
section~\ref{sec_skob_alpha} seems to be promising \cite{EissWag}.

%%%%%%%%%%%%%%%%%%%%%%%%%%%%%%%%%%%%%%%%%%%%%%%%%%%%%%%%%%%%%%%%%%%%%%%%%%%%%
%
\paragraph{Acknowledgments}

We like to thank Robert Barlovic, Kai Nagel and Gunter Sch\"utz for useful 
discussions.
AS thanks the DAAD for partial support and the Institute for Mathematics
and Statistics at the University of Sao Paulo for its hospitality during
the final stages of this work.
%%%%%%%%%%%%%%%%%%%%%%%%%%%%%%%%%%%%%%%%%%%%%%%%%%%%%%%%%%%%%%%%%%%%%%%%%%%%%
%%%%%%%%%%%%%%%%%%%%%%%%%%%%%%%%%%%%%%%%%%%%%%%%%%%%%%%%%%%%%%%%%%%%%%%%%%%%%
%
% ---- Bibliography ----
%

%%
%%
\end{document}